\documentstyle[a4,psfig]{article}

\begin{document}

\title{Phase equivalent potentials for three-body halos}
\author{E. Garrido  \\ Instituto de Estructura de la Materia,
CSIC, \\ Serrano 123, E-28006 Madrid, Spain \\ ~\\ and \\ ~\\ 
D.V.~Fedorov and A.S.~Jensen \\ 
Institute of Physics and Astronomy, \\
University of Aarhus, DK-8000 Aarhus C, Denmark }

\maketitle

\begin{abstract}
We compare the properties of three-body systems obtained with two-body
potentials with Pauli forbidden states and with the corresponding
phase equivalent two-body potentials. In the first case the forbidden
states are explicitly excluded in the calculation. Differences arise
due to the off-shell properties of these on-shell equivalent
potentials. We use the adiabatic hyperspherical method to formulate a
practical prescription to exclude Pauli forbidden states in three-body
calculations. Schematic as well as realistic potentials are
used. Almost indistinguishable results are obtained. \\

\noindent
PACS number(s): 21.60.Gx, 21.45.+v  
\end{abstract}


\section{Introduction}

Cluster models are useful to describe spatially correlated systems
when the intrinsic cluster structures essentially are frozen and the
relative and intrinsic degrees of freedom essentially are decoupled.
A particular application during the last decade has been in studies of
nuclear halos \cite{han95,jon98} and the concepts developed in this
connection have already penetrated in to other subfields of physics
\cite{esr96,nie97,nie98}.

When different clusters contain identical particles the Pauli
forbidden states must often be excluded. This is conceptually simple
but often difficult in practise, since the proper antisymmetrization
too easily spoils the advantages of the cluster description.  For
spatially extended systems with little overlap the details of
appropriate approximations have so far been rather unimportant
\cite{zhu93,gar97,joh90,rid96,cob98,gar98}.  However, beside the
conceptual interest better treatments may be required by the need for
higher accuracy.

Furthermore, the validity of the cluster models may be extended to
less pronounced halos by an accurate treatment of the exclusion
principle. The dilemma is that the cluster models become more accurate
with decreasing overlap while the overlap is necessary to provide the
interesting structure and the binding energy. Too large an overlap
destroys the fundamental decoupling of the intrinsic and relative
cluster degrees of freedom, but the exclusion principle becomes
important already at a smaller density overlap.

The correlations displayed in two and three-body halos have attracted
a lot of interest and in particular those in light nuclei. If the halo
consists of one nucleon bound to a nucleus the Pauli forbidden states
are easily excluded by precise orthogonalization. However, already
three-body systems present more difficulties and approximations are
often employed. Various prescriptions have so far been used in
numerical calculations on halos consisting of two nucleons and a core
\cite{zhu93,gar97}.  The simplest is to use a nucleon-core effective
potential without bound states \cite{joh90} even when the
nucleon then moves in a potential different from that of the identical
nucleons within the core. The difference amounts in this way to an
additional repulsion accounting for the Pauli exclusion principle.

It is essential that this effective potential reproduces the
nucleon-core low-energy scattering properties, i.e. scattering length
and effective range or even better the phase shifts as function of
energy. Therefore an obvious procedure \cite{and84,suk85} is to start
with a deep potential appropriate for the nucleons in the core. Then
construct a phase equivalent, approximate or precise, shallow or
repulsive potential \cite{bay87} and compute the three-body
observables of interest.  The results could be compared to those
obtained with the deep potential and the Pauli forbidden states
excluded in one way or another \cite{gar97,fie90}.

This procedure must produce two sets of wave functions coinciding at
large distances but differing at small distances by the additional
node appearing inside the core by use of the deep potential. The idea
is therefore that the short-distance behavior must be irrelevant when
the cluster model provides a valid description. This conjecture was
tested in three-body computations by using two potentials with the
same scattering length and effective range \cite{gar97} and in
two-body computations using completely phase equivalent potentials
\cite{rid96}.

The purpose of this paper is to investigate the consequences of using
completely phase equivalent two-body potentials
\cite{rid96,and84,suk85,bay87,fie90} in three-body calculations based
on the adiabatic hyperspherical approach, see
e.g. \cite{esr96,nie98}. The method was suggested and previously
tested approximately for two neutrons and a core \cite{gar97}.  In
this report we shall carefully follow this prescription using phase
equivalent potentials and rigorously derive the short and
long-distance behavior of the all-decisive effective radial
potentials. We shall consider cases when one, two or three of the
two-body potentials have Pauli forbidden states characterized by
various quantum numbers.

We begin with a brief sketch of the general method in section 2. We
then in section 3 give the characteristic properties of the
constructed phase equivalent two-body potentials, where we use a
simple $s$-wave potential as an analytical example. In section 4 we
derive the short and long-distance properties of the three-body
adiabatic effective radial potentials. In section 5 we show a number
of qualitatively different numerical examples. Finally section 6
contains a summary and the conclusions.

\section{Adiabatic hyperspherical expansion}

In the adiabatic hyperspherical method the coordinates of the
three-body system are divided into the slow (hyperradius) and fast
(angular) variables.  The eigenvalues and eigenfunctions for the fast
angular degrees of freedom are then calculated for fixed values of the
slow radial variable. These eigenfunctions are then used as a basis
for expansion of the total wave function. Eigenvalues for the fast
angular variables serve as adiabatic potentials for the slow radial
variable. The great advantage of this basis is that it correctly
reproduces the long-range asymptotic behavior of the three-body
systems including the pathological Efimov effect.

\subsection{Hyperspherical coordinates}

We first introduce the usual Jacobi coordinates \cite{zhu93} for three
particles with masses $m_{i}$ and coordinates ${\bf r}_{i}$:
\begin{equation}
{\bf x}_{i}=\sqrt{\frac{1}{m}\frac{m_{j}m_{k}}{m_{j}+m_{k}}}
({\bf r}_{j}-{\bf r}_{k})\;,\;
{\bf y}_{i}=\sqrt{\frac{1}{m}\frac{m_{i}(m_{j}+m_{k})}{M}}
\left( {\bf r}_{i}-\frac{m_{j}{\bf r}_{j}+m_{k}{\bf r}_{k}}
{m_{j}+m_{k}} \right) \;, 
\end{equation}
where $M=m_1+m_2+m_3$ and $m$ is an arbitrary mass scale. The
hyperspherical coordinates now consist of a hyperradius $\rho
=\sqrt{x_{i}^{2}+y_{i}^{2}}$ and five dimensionless angles $\Omega
=\{\alpha_i,\Omega_{xi},\Omega_{yi}\}$, where $\Omega_{xi}$ and
$\Omega_{yi}$ specify each of the directions of ${\bf x}_{i}$ and
${\bf y}_{i}$ and $\alpha_i \equiv \arctan (x_{i}/y_{i})$.  The
kinetic energy operator $T$ can be written in terms of the
hyperspherical coordinates as
\begin{eqnarray}
T &=&T_{\rho }+\frac{\hbar ^{2}}{2m\rho ^{2}}\widehat{\Lambda}_{\Omega }^{2} 
\;, \\
T_{\rho } &=&-\frac{\hbar ^{2}}{2m}\rho ^{-5/2}\frac{\partial ^{2}}{\partial
\rho ^{2}}\rho ^{5/2}+\frac{15}{4}\frac{\hbar ^{2}}{2m\rho ^{2}} \; , \\
\widehat{\Lambda}_{\Omega }^{2} &=&-\frac{1}{\sin (2\alpha )}\frac{\partial ^{2}}{%
\partial \alpha ^{2}}\sin (2\alpha )-4+\frac{\widehat{l}_{x}^{2}}{\sin
^{2}\alpha }+\frac{\widehat{l}_{y}^{2}}{\cos ^{2}\alpha }\;,
\end{eqnarray}
where $\widehat{l}_{x}^{2}$ and $\widehat{l}_{y}^{2}$ are the angular
momentum operators related to the ${\bf x}$ and ${\bf y}$ coordinates.

\subsection{Adiabatic expansion}

The three-body hamiltonian is given by
\begin{equation}
H=T_{\rho }+\frac{\hbar ^{2}}{2m\rho ^{2}}\widehat{\Lambda}_{\Omega
}^{2}+\sum_{i}V_{i}(\rho ,\Omega ) \; ,
\end{equation}
where $V_i$ is the interaction between the two particles $j$ and
$k$. Here \{$i,j,k$\} is a permutation of \{1,2,3\}. The eigenvalue
problem is solved in two steps.  First the slow coordinate $\rho $ is
fixed and the (always discrete) spectrum $\lambda _{n}(\rho )$ and the
corresponding angular eigenfunctions $\Phi _{n}(\rho ,\Omega )$ are
calculated as function of $\rho$. For this one can either use the
Schr\"{o}dinger equation
\begin{equation} \label{angular}
\left( \widehat{\Lambda}_{\Omega }^{2}+\frac{2m\rho ^{2}}{\hbar ^{2}}%
\sum_{i}V_{i}-\lambda _{n}(\rho )\right) \Phi _{n}(\rho ,\Omega )=0\;,
\end{equation}
or the equivalent Faddeev equations 
\begin{eqnarray} \label{angfad}
 \left( \widehat{\Lambda}_{\Omega }^{2}-\lambda _{n}(\rho )\right) 
 \frac{\phi _{n}^{(i)}(\rho,\Omega )}{\sin(2 \alpha_i)}
 + \frac{2m\rho ^{2}}{\hbar ^{2}}V_{i}
 \;\Phi _{n}(\rho ,\Omega ) &=&0\;,\;i=1,2,3 \;,
\end{eqnarray}
where the Schr\"{o}dinger wave function $\Phi _{n}$ is the sum of the
Faddeev components $\phi _{n}^{(i)}$ or more precisely $\Phi _{n}(\rho
,\Omega) = \sum_{i}\phi _{n}^{(i)}(\rho ,\Omega) / \sin(2 \alpha_i)$.

The total wave function $\Psi (\rho ,\Omega)$ is expanded on the
complete angular basis set $\Phi _{n}$, i.e.
\begin{equation}
 \Psi(\rho ,\Omega )= \frac{1}{\rho^{5/2}}
 \sum_{n}f_{n}(\rho )\Phi _{n}(\rho ,\Omega) =  \frac{1}{\rho^{5/2}}
 \sum_{ni} f_{n}(\rho ) \frac{\phi _{n}^{(i)}(\rho ,\Omega)}{\sin(2 \alpha_i)}
 \; ,
\end{equation}
where the phase space factors $\rho^{-5/2}$ and $\left[\sin(2
\alpha_i)\right]^{-1}$ for convenience are extracted explicitly.

The radial expansion coefficients $f_{n}(\rho )$ are then solutions to
\begin{eqnarray}
&&\left( -\frac{\hbar ^{2}}{2m}\frac{\partial ^{2}}{\partial \rho ^{2}}+
\frac{\hbar ^{2}}{2m\rho ^{2}}\left( \lambda _{n}(\rho )+\frac{15}{4}\right)
-Q_{nn}(\rho ) + V_3(\rho) -E\right) f_{n}(\rho ) \nonumber \\ \label{rad1}
&=&\sum_{n^{\prime }\neq n}\left( -2P_{nn^{\prime }}(\rho )\frac{\partial }{
\partial \rho }-Q_{nn^{\prime }}(\rho )\right) f_{n^{\prime }}(\rho )\;,
\end{eqnarray}
where $V_3(\rho)$ is an anticipated three-body potential for fine
tuning and the coupling terms $P$ and $Q$ are given as
\begin{eqnarray}
P_{nn^{\prime }}(\rho ) &=&\int d\Omega \;\Phi _{n}^{*}(\rho ,\Omega )
\frac{\partial }{\partial \rho }\Phi _{n^{\prime }}(\rho ,\Omega ) \; ,\\
Q_{nn^{\prime }}(\rho ) &=&\int d\Omega \;\Phi _{n}^{*}(\rho ,\Omega )
\frac{\partial ^{2}}{\partial \rho ^{2}}\Phi _{n^{\prime }}(\rho ,\Omega )\;
\end{eqnarray}
with the angular volume element defined by ${\rm d}\Omega=\sin^2
\alpha \cos^2 \alpha {\rm d}\alpha {\rm d}\Omega_x {\rm d}\Omega_y$.
When all but the lowest terms are neglected we obtain the usual
adiabatic approximation.

The angular eigenvalues $\lambda _{n}(\rho )$ are therefore closely
related to the adiabatic radial potentials. The behavior of these
angular eigenvalues decisively determines the properties of the
three-body system.  We shall therefore analytically and numerically
investigate $\lambda _{n}(\rho )$ for phase equivalent two-body
potentials differing by the number of bound states. We shall
especially study the asymptotic properties analytically.

\section{Phase equivalent two-body potentials}

Let us assume that a two-body system with relative coordinate ${\bf
r}$, reduced mass $\mu$ and potential $V$ has at least one bound state
with the energy $E_{lsj} = -\hbar^{2}\kappa ^{2}/2\mu $ and the total
wave function $\psi_{lsj}({\bf r}) = u_{lsj}(r) r^{-1} {\cal
Y}_{lsjm}$ expressed in terms of the radial wave function $u_{lsj}(r)$
and the spin-angular part ${\cal Y}_{lsjm}$, where $(s,l,j)$ are the
spin, the orbital and total angular momenta, respectively, while $m$
here is the projection of ${\bf j}$ on the $z$-axis. We then construct
a new potential
\begin{eqnarray} \label{pep1}
V_{pep}({\bf r}) &=&V({\bf r}) + \Delta V_{lsj}(r) \;, \\ \label{pep2}
\Delta V_{lsj}(r) &=& - \widehat {P}_{j} \widehat {P}_{l} \widehat {P}_{s}
\frac{\hbar ^{2}}{\mu }\frac{d^{2}}{dr^{2}}\ln \left(
\int_{0}^{r}\left|u_{lsj}(r^{\prime })\right|^{2}dr^{\prime }\right) \; ,
\end{eqnarray}
where $(\widehat {P}_j,\widehat {P}_l,\widehat {P}_s)$ are projection
operators on states with the corresponding angular momentum quantum
numbers. Thus the additional potential $\Delta V_{lsj}(r)$ is an
effective radial potential which only is active for the spin and
angular momentum quantum numbers of the bound state in question. The
potential $V_{pep}({\bf r})$ is only missing the bound state
$\psi_{lsj}$ of $V({\bf
r})$, but otherwise these two potentials have identical properties,
i.e. they are phase equivalent with the same bound state spectrum
\cite{rid96,and84,suk85,bay87,fie90}.

If desired this construction can be repeated to eliminate possible
other bound states of the new potential. The series of potentials then
only differ by the number of bound states. 

The wave function $u_{lsj}(r)$ has the following asymptotic forms at
short and large distances
\begin{equation}
 u_{lsj}(r\rightarrow 0)\propto r^{l+1} \;\;\; , \;\;\;
 u_{lsj}(r\rightarrow \infty ) \propto  e^{-\kappa r} \; .
\end{equation}
The additional term $\Delta V_{lsj}$ in Eq.(\ref{pep1}) diverges as
$r^{-2}$ at small distances and vanishes exponentially at large
distances \cite{rid96,and84,suk85,bay87,fie90}
\begin{eqnarray} \label{dpo1}
\Delta V_{lsj}(r\rightarrow 0)\rightarrow 
\widehat {P}_{j} \widehat {P}_{l} \widehat {P}_{s} \frac{\hbar ^{2}}{\mu }
\frac{2l+3}{r^{2}}\;  , \\ \label{dpo2}
\Delta V_{lsj}(r\rightarrow \infty )
  \rightarrow \widehat {P}_{j} \widehat {P}_{l} \widehat {P}_{s}
 \frac{4 \hbar ^{2} \kappa^{2}}{\mu } \exp (-2\kappa r)\;.
\end{eqnarray}

We shall in the remaining part of the present paper often consider an
original potential $V$ with one bound $s$-state. This is the most
important case in the practical applications for halo states and a
generalization is easy and straightforward by use of Eqs.(\ref{pep1})
and (\ref{pep2}). Properties and procedure can be studied in more
details in simple cases. For illustration we therefore construct an
analytical $s$-wave potential and the corresponding solution labeled
by ``0''. The wave function $u_{0}(r)$ must be proportional to $r$ at
small distances and must fall off exponentially at large
distances. One of the simplest functions obeying these constraints is
\begin{equation} \label{wav0}
u_{0}(r)=\sinh (\beta r)\exp (-\alpha r) \; ,\
\end{equation}
which is the solution with the energy $E_0$ for the potential $V_0$
given by
\begin{equation} \label{pep0}
 E_{0} = -\frac{\hbar ^{2}}{2\mu }(\alpha -\beta )^{2} \equiv
 - \frac{ \hbar ^{2} \kappa^{2}}{2\mu } \;\;,\;\;
 V_0(r)=\frac{\hbar^{2}}{2\mu }2\beta \alpha \left[1-\coth (\beta r)\right]\;,
\end{equation}
where $\kappa = \alpha - \beta > 0$.

The asymptotic behavior of this potential is then
\begin{eqnarray} \label{asy1}
  V_0(r \rightarrow 0)=  \frac{\hbar ^{2}}{2\mu} \left( - \frac{2\alpha}{r}
 + 2\beta \alpha + O(r) \right) \;, \;   
   V_0(r\rightarrow \infty ) =  - \frac{2\hbar ^{2}}{\mu}
 \beta \alpha \exp (-2\beta r) \; , \label{asy2}
\end{eqnarray}
where $O(r)$ is a function vanishing at least as fast as $r$ when $r$
decreases towards zero.

The phase equivalent potential in Eq.(\ref{pep1}) has the additional
term $\Delta V_0$ defined in Eq.(\ref{pep2}). For the example in
Eqs.(\ref{wav0}) and (\ref{pep0}) we then get
\begin{eqnarray}
\Delta V_0(r) = \frac{\hbar ^{2}}{\mu }8\alpha \beta ^{2}(\alpha ^{2}-\beta
^{2})\sinh (\beta r) \nonumber \\ \times
\frac{\alpha (e^{2\alpha r}+1)\sinh (\beta r)-\beta
(e^{2\alpha r}-1)\cosh (\beta r)}{\left( \alpha ^{2}\left( \cosh (2\beta
r)-1\right) -\beta ^{2}\left( e^{2\alpha r}-1\right) +\alpha \beta \sinh
(2\beta r)\right) ^{2}} = \nonumber \\ 
\frac{\hbar ^{2}}{\mu }\frac{16\alpha \beta ^{2}(\alpha ^{2}-\beta
^{2})e^{\alpha r}\sinh (\beta r)\left[ \alpha \cosh (\alpha r)\sinh (\beta
r)-\beta \sinh (\alpha r)\cosh (\beta r)\right] }{\left[ \alpha ^{2}\left(
\cosh (2\beta r)-1\right) -\beta ^{2}\left( e^{2\alpha r}-1\right) +\alpha
\beta \sinh (2\beta r)\right] ^{2}} \; , \label{dpep0}
\end{eqnarray}
which only acts on $s$-waves and asymptotically behaves in agreement
with Eqs.(\ref{dpo1}) and (\ref{dpo2}).

\section{Asymptotics of the angular eigenvalues}

The different two-body potentials from section 3 should now be used in
the three-body computations described in section 2. Two-body
interactions are needed for the three two-body subsystems. Each of
these can in principle support Pauli forbidden states of specific but
different properties. In any case the angular eigenvalue spectrum, or
equivalently the adiabatic radial potentials, is crucial for the
behavior of the resulting three-body system. We shall therefore
calculate, as far as possible analytically, the leading terms in the
asymptotic expansion of the adiabatic radial potentials both for small
and large $\rho $. Numerical examples for all $\rho$-values are
postponed to the next section.

\subsection{Short distances}

Let us assume that the original two-body potentials are of short range
and finite at $r=0$. The different angular momentum states then
decouple in the three-body Schr\"{o}dinger equation in the
small-distance limit $\rho = 0$. For each set of quantum numbers $l_x$
and $l_y$ referring to the $i'$th set of Jacobi coordinates, the
angular eigenvalue equation in Eq.(\ref{angular}) reduces to
\begin{eqnarray} 
\left( - \frac{1}{\sin (2\alpha_i)} \frac{\partial ^{2}}{\partial \alpha ^{2}}
  \sin (2\alpha_i) +
 \frac{l_x(l_x+1)}{\sin ^{2}\alpha_i } + \frac{l_y(l_y+1)}{\cos ^{2}\alpha_i }
 \right. \nonumber \\  \left.
+ \frac{2m\rho ^{2}}{\hbar ^{2}}\sum_{j}V_{j}\right) 
 \Phi _{nl_xl_y}=\nu _{n}^{2}\Phi _{nl_xl_y} \; , \label{schr1}
\end{eqnarray}
where $\lambda _{n}=\nu _{n}^{2}-4$ and the potentials are kept for
later use although their contributions vanish for $\rho =0$. The
eigenvalues and the corresponding eigenfunctions are
\cite{zhu93,cob98,jen97}
\begin{eqnarray} \label{eig1}   
 &  \nu _{n}=2(n+1) \;\;\; , \;\;\;  n=0,1,2,...  \\ \label{fct1}   
&  \Phi_{nl_xl_y} =  \frac{1}{\sin (2\alpha_i)} P_{n}^{(l_x+1/2,l_y+1/2)}
 (\cos 2\alpha_i ) Y_{l_xm_x}(\Omega_{xi}) Y_{l_ym_y}(\Omega_{yi}) \; ,
\end{eqnarray}
where $Y_{lm}$ and $P_{n}^{(l_x+1/2,l+y1/2)}$ are the spherical
harmonics and the associated Legendre polynomials, respectively.

Let us now assume that only one bound state of angular momentum $l_x$
is Pauli forbidden in one of the two-body subsystems, which we choose
to relate to the $x$-coordinate and denote by $i$.  The phase
equivalent potential in Eq.(\ref{pep1}) then has an additional term
$\Delta V$, which for small $\rho $ contributes to Eq.(\ref{schr1}) as
\begin{equation}
 \frac{2m\rho ^{2}}{\hbar ^{2}}\Delta V \;  \widehat {P}_{l_x} \approx
 \frac{2m\rho ^{2}}{\hbar ^{2}} \frac{(3+2l_x)\hbar ^{2}}{\mu r^{2}} 
\widehat {P}_{l_x} = \frac{2(3+2l_x)}{\sin ^{2}\alpha_i } 
 \widehat {P}_{l_x} \;,
\end{equation}
where we used that $\mu r^{2}/m=\rho ^{2}\sin ^{2}\alpha_i $. All other
partial waves remain unchanged.

The Schr\"{o}dinger equation in Eq.(\ref{schr1}) is then for this
particular $l_x$ modified at small $\rho$ into
\begin{eqnarray}
 \left( - \frac{1}{\sin (2\alpha)} \frac{\partial ^{2}}
 {\partial \alpha^{2}} \sin (2\alpha)  +
 \frac{(l_x+2)(l_x+3)}{\sin ^{2}\alpha } + 
 \frac{l_y(l_y+1)}{\cos ^{2}\alpha }  \right. \nonumber \\  \left.
 + \frac{2m\rho ^{2}}{\hbar ^{2}}\sum_{j}V_{j}\right) \widetilde{
\Phi }_{nl_xl_y}=\widetilde{\nu }_{n}^{2}\widetilde{\Phi }_{nl_xl_y} \; ,
 \label{schr2}
\end{eqnarray}
where we omitted the index $i$ on the coordinates and tildes indicate
the new eigenvalues and eigenfunctions. Thus the Schr\"{o}dinger
equation is changed as if the angular momentum quantum number $l_x$
were increased by two units. For all other partial waves with $l_x$
different from the quantum number of the two-body bound state the
Schr\"{o}dinger equation remains unchanged.  The new spectrum and the
corresponding eigenfunctions are given by 
\begin{eqnarray} \label{fct3}   
 & \widetilde{\nu} _{n}=2(n+2) \;\;\; , \;\;\;  n=0,1,2,...  \\ \label{fct2}   
&  \widetilde{\Phi}_{nl_xl_y} =  \frac{1}{\sin (2\alpha)} 
 P_{n}^{(l_x+5/2,l_y+1/2)}
 (\cos 2\alpha ) Y_{l_xm_x}(\Omega_x) Y_{l_ym_y}(\Omega_y) \; .
\end{eqnarray}
Thus, except for the missing lowest eigenvalue, the three-body angular
spectrum for the phase equivalent potential is identical to the
original potential up to the first two terms in the expansion for
small $\rho $, i.e.
\begin{equation}\label{small}
\nu _{n}^{2}=(2(n+1))^{2}+\frac{2m\rho ^{2}}{\hbar^{2}}\sum_{i}V_{i}(0) \;\;\;,
 \;\;\; \widetilde{\nu }_{n}^{2}=(2(n+2))^{2}+\frac{2m\rho ^{2}}{\hbar ^{2}}
\sum_{i}V_{i}(0) \; ,
\end{equation}
where $n=0,1,2,...$.

The next step is to include more than one two-body bound state in the
same subsystem. If the orbital angular momenta of these states are
different the three-body angular equations decouple and each of them
is analogous or identical to Eq.(\ref{schr2}). Thus each Pauli
forbidden state is removed independently. On the other hand if the
angular momenta are equal the procedure for one two-body bound state
is repeated and two units have to be added to the corresponding $l_x$
both in Eq.(\ref{schr2}) and in the Legendre polynomial of
Eq.(\ref{fct2}), i.e. $P_{n}^{(l_x+5/2,l_y+1/2)}$ should be
substituted by $P_{n}^{(l_x+9/2,l_y+1/2)}$ while the spherical
harmonics remain unchanged. The spectrum is again identical to that of
the original potential except for the missing two lowest eigenvalues.

Let us now assume that more than one of the two-body subsystems have
Pauli forbidden states. Using the phase equivalent potentials the
three-body Schr\"{o}dinger equation for small $\rho$ then becomes
\begin{eqnarray} 
 \left( - \frac{1}{\sin (2\alpha_i)} 
\frac{\partial ^{2}}{\partial \alpha_i ^{2}} \sin (2\alpha_i)   +
 \frac{ \widehat{l}_{x}^{2}}{\sin ^{2}\alpha_i } 
 +  \frac{ \widehat{l}_{y}^{2}}{\cos ^{2}\alpha_i }  
+ \sum_{j} \frac{2(3+2l_{xj}^{(0)})}{\sin ^{2}\alpha_j } 
 \widehat {P}_{l_{xj}^{(0)}} \right. \nonumber \\  \left.
 + \frac{2m\rho ^{2}}{\hbar ^{2}}\sum_{j}V_{j}\right) \widetilde{
\Phi }_{nL}= 
  \widetilde{\nu }_{n}^{2}\widetilde{\Phi }_{nL} \; , \label{schr3} 
\end{eqnarray}
where $j$ in the first summation includes subsystems with a Pauli
forbidden bound state of angular momentum $l_{xj}^{(0)}$. The
projection operator projects on states with the angular momentum
$l_{x}=l_{xj}^{(0)}$ in the $j'$th Jacobi coordinate system.  When more
than one subsystem has a Pauli forbidden state the different partial
waves couple already in the limit $\rho = 0$. Therefore we only
maintained the conserved total angular momentum $L$ as an index on the
solutions.

The spectrum arising from Eq.(\ref{schr3}) can not be obtained
analytically. Let us then first repeate that treating each of the
additional potentials $\Delta V$ while neglecting the others results
in Eq.(\ref{fct2}). Thus simultaneous inclusion of more than one of
these potentials related to different subsystems could be expected to
move the spectrum by approximately the sum of that obtained by each of
the individual treatments. The corresponding approximate wave
functions are analogously obtained from Eq.(\ref{fct2}) with the
appropriate addition of units to the angular momentum quantum numbers
in the Legendre polynomials. However, this is not the exact solution
due to the coupling of different partial waves. If the states with
$l_{x1}^{(0)}$ and $l_{x2}^{(0)}$ are Pauli forbidden in the
subsystems labeled ``1'' and ``2'', respectively, all $l_{x1}$ in
subsystem ``1'' couple provided they receive finite contributions from
rotation of the state with $l_{x2}^{(0)}$. Thus numerical solutions
are required to get the precise solutions.

A better, still approximate, analytically obtained spectrum can be
found by first solving the equation when $l_{x1}$ differs from all
three $l_{xj}^{(0)}$ and all additional $\Delta V$ terms are
neglected. The wave functions are then the usual hyperspherical
harmonics. Then when $l_{x1}$ equals one and only one of the values
$l_{xj}^{(0)}$ we use the wave functions in Eq.(\ref{fct2}). When
$l_{x1}$ is equal to two or three identical $l_{xj}^{(0)}$-values
the Legendre polynomial in Eq.(\ref{fct2}) has to be substituted by
$P_{n}^{(l_x+9/2,l_y+1/2)}$ or $P_{n}^{(l_x+13/2,l_y+1/2)}$,
respectively.  The spectrum is finally found approximately by using
these approximate wave functions $\widetilde{\Phi}_{nl_xl_y}^{(a)}$
and first order perturbation $\Delta E_{nl_xl_y}$ on top of the
hyperspherical spectrum where one or more of the lowest levels are
missing. The energy correction is then
\begin{equation} \label{per}
 \Delta E_{nl_xl_y} = \sum_{j} \langle \widetilde{\Phi}_{nl_xl_y}^{(a)} \Big|
 \frac{2(3+2l_{xj}^{(0)})}{\sin ^{2}\alpha_j } \widehat {P}_{l_{xj}^{(0)}}
 - \frac{2(3+2l_{xj}^{(0)})}{\sin ^{2}\alpha_1} 
 \widehat {P}_{l_{x1}=l_{xj}^{(0)}}  \Big| 
\widetilde{\Phi}_{nl_xl_y}^{(a)} \rangle \; ,
\end{equation}
where again the summation is over the Pauli forbidden states. This
means that at most two non-zero terms are present. Unfortunately it
not easier to use this approximation than to compute the spectrum
numerically without any approximations.

In conclusion, in the limit of small $\rho$ the original and the phase
equivalent potentials have, except for the Pauli forbidden
eigenvalues, exactly the same spectrum of angular eigenvalues when all
Pauli forbidden states occur in one two-body subsystem. When more than
one subsystem support Pauli forbidden states the spectrum is only
approximately the same for $\rho = 0$, again excluding the
lowest-lying Pauli forbidden levels.

\subsection{Large distances}

The hyperspherical spectrum is for short-range two-body potentials
also obtained in the limit of $\rho = \infty$ as established
previously \cite{nie98,jen97,fed93,fed94}. It is essentially
equivalent to the spectrum at $\rho = 0$ given by Eq.(\ref{eig1}).
However, in addition some of the angular eigenvalues diverge
parabolically towards $-\infty$ in a one-to-one correspondence with
the bound states in the two-body subsystems. The corresponding wave
functions are the hyperspherical harmonics in Eq.(\ref{fct1}) except
for the lowest diverging levels where the structure at $\rho = \infty$
is that of two particles in a bound state while the third particle is
in a continuum state relative to this two-body structure \cite{fed94}.

The convergence towards the hyperspherical spectrum is faster the
higher the angular momentum quantum numbers. The limits are reached at
distances comparable to the effective ranges of the two-body
interactions \cite{nie98,cob98} with the crucial exception of
$s$-waves, where the two-body scattering lengths define the distance
for convergence \cite{jen97}. The different partial waves decouple in
the sense that contributions to a given state from higher angular
momenta are at least one order smaller in expansions in terms of
$\rho^{-1}$. Thus for example $p$-waves only receive contributions of
the same order in the $\rho^{-1}$ expansion from the $s$-waves, which
in turn only receive contributions of the same order from other
$s$-waves. These couplings are basically determined by the centrifugal
barriers increasing with angular momentum.

The relatively fast convergence for the higher angular momentum states
isolate the $s$-waves as the only interesting contributors at large
distances. The diverging levels are of special interest, since some of
them correspond to Pauli forbidden states, which must disappear by
changing to the phase equivalent potentials in Eq.(\ref{pep1}).  In
fact the goal of the present subsection is to compare the three-body
angular eigenvalues of deep and phase equivalent two-body potentials
differing only by the number of bound states. This should then allow
us to design a prescription for removing Pauli forbidden states at
large distance.

The necessary subsequent connection from small to large distances
requires knowledge of the approach to the asymptotic behavior, i.e. at
least a minimum of information about rate of convergence and wave
function decomposition during the approach. Here the $s$-waves are
crucial and we shall therefore derive a number of characteristic
features at large distances, i.e. corresponding to distances $\rho$
simultaneously outside the effective ranges of all the two-body
interactions. These properties are revealed by use of the zero-range
potentials with the related analytical solutions.

For short-range potentials and large distances the Faddeev equations
are more convenient than the Schr\"{o}dinger equation.  We therefore
start with Eq.(\ref{angfad}) for $s$-waves, which means that the
partial angular momenta $l_x = l_y = 0$ in all three Jacobi
coordinates. We choose the system labeled $i$ and transform the other
two Faddeev components into this system, project out the $s$-wave
components and multiply by $\sin(2 \alpha_i)$. The wave function
resulting from $\Phi _{n}$ after these operations is denoted by $\Phi
_{n}^{(i)}$ and from Eq.(\ref{angfad}) we then obtain the equation of
motion as
\begin{eqnarray}
\left( -\frac{\partial ^{2}}{\partial \alpha_i ^{2}}-\nu _{n}^{2}\right) \phi
_{n}^{(i)}(\alpha _{i})+\frac{2m\rho ^{2}}{\hbar ^{2}}V_{i}(\sqrt{\frac{\mu
_{i}}{m}}\rho \sin \alpha _{i})\Phi _{n}^{(i)}(\alpha _{i}) = 0\;,\;i=1,2,3,
\label{fad1} \\
\Phi _{n}^{(i)}(\alpha ) = \phi _{n}^{(i)}(\alpha )+\sum_{j\neq i}\frac{1}{%
\sin (2\varphi _{ij})}\int_{\left| \varphi _{ij}-\alpha \right| }^{\frac{\pi 
}{2}-\left| \frac{\pi }{2}-\varphi _{ij}-\alpha \right| }d\alpha ^{\prime
}\phi _{n}^{(j)}(\alpha ^{\prime })\;, \label{fad2}
\end{eqnarray}
where $\varphi _{ij}=(-1)^{P(ijk)}\arctan \left(
\sqrt{m_{k}(m_{1}+m_{2}+m_{3})/(m_{i}m_{j})} \right) $ and $P(ijk)$
is the parity of the permutation $\{ijk\}$ while $m$ and $\mu
_{i}=m_{j}m_{k}/(m_{j}+m_{k})$ are the normalization mass $m$ and the
reduced mass, respectively. The last term in Eq.(\ref{fad2}) arises
from the transformation of the two Faddeev components different from
$i$, see for example \cite{nie98,zhu93,jen97}. A complete solution for
square well potentials can be found in \cite{jen97}. 

For large distances and short-range potentials $V_i$ vanish when
$\alpha_i$ is much larger than the largest effective range of the
potentials divided by $\rho$. In this region the free solution is
obtained for each Faddeev component $\phi _{n}^{(i)}$. In the
remaining vanishingly small region of $\alpha_i$-space around
$\alpha_i = 0$ the potential $V_i$ contributes.  In the region of
small $\alpha_i$ the integral in the function $\Phi _{n}^{(i)}$ in
Eq.(\ref{fad2}) can be expanded in terms of $\alpha_i$. To first order
we then obtain
\begin{equation} \label{limit}
\Phi _{n}^{(i)}(\alpha \ll 1)=\phi _{n}^{(i)}(\alpha )+2\alpha \sum_{j\neq i}%
\frac{\phi _{n}^{(j)}(\varphi _{ij})}{\sin (2\varphi _{ij})}\;,\;i=1,2,3\;.
\end{equation}
For large distances $\rho \gg R \equiv \max (R_{1},R_{2},R_{3})$,
where $R_{i}$ is the range of the potential $V_{i}$, the solution to
Eq.(\ref{fad1}) in the $\alpha_i$-region, $x_i \equiv \sqrt{\frac{\mu
_{i}}{m}} \rho \sin \alpha _{i} > R$, is the free solution
\begin{equation} \label{sin}
\;\phi ^{(j)}(\alpha)=A_{j}\sin \left[ \nu \left( \alpha -\frac{\pi }{%
2}\right) \right] \;,  
\end{equation}
where $A_{j}$ is constant. The structure of this solution is by
definition independent of the potentials. The wave function $\phi
_{n}^{(j)}(\varphi _{ij})$ in Eq.(\ref{limit}) is then given by 
Eq.(\ref{sin}), since the argument $\varphi _{ij}$ is finite.

Increasing the parameter $\rho $ in Eq.(\ref{fad1}) effectively
decreases the range and increases the strength of the potential and
thus the zero-range limit is approached. The related eigenvalue
problem can then be reformulated in the limit $\rho \rightarrow \infty
$ for zero-range potentials. For $x_i< R_i$ and $\alpha_i \ll 1$
Eq.(\ref{fad1}) is approximately given as
\begin{eqnarray} \label{fad3}
 \left( - \frac{\hbar ^{2}}{2\mu_{i}} \frac{\partial ^{2}}{\partial x_i ^{2}}
 - \frac{\hbar ^{2}\nu _{n}^{2}}{2 m \rho^2} \right) 
 \phi_{n}^{(i)}(x _{i}) + V_{i}(x_i) \Phi _{n}^{(i)}(x_{i}) = 0
 \;\;\;,\;\;\; i=1,2,3
\end{eqnarray}
and for $x_i > R_i$ the potential vanishes. The Schr\"{o}dinger
equation for $\Phi _{n}^{(i)}$ then arises by adding the three
equations in Eq.(\ref{fad3}).

The singular points are $\alpha_i =0$ or $x_i=0$ and the boundary
conditions for these zero-range potentials are then expressed in terms
of the two-body scattering lengths $a_{i}$ as
\begin{equation} \label{bou}
 \left[ \frac{1}{\Phi_{n}^{(i)}} \frac{\partial }{\partial x_i }
 \Phi _{n}^{(i)}\right] _{x_i =0} = \frac{1 }{{a}_{i}} \;\;\; , \;\;\;
\left[ \frac{1}{\Phi_{n}^{(i)}} \frac{\partial }{\partial \alpha_i }
 \Phi _{n}^{(i)}  \right] _{\alpha_i =0}
 = \frac{\rho }{\widetilde{a}_{i}}  
 \;\; , \;\;\; i=1,2,3 \;,
\end{equation}
where we define the modified scattering lengths by $\widetilde{a}_{i}
= a_{i} \sqrt{\mu_i/m}$. Thus in the zero-range limit the eigenvalue
spectrum is determined entirely by the scattering lengths and
identical spectra results from potentials with the same scattering
lengths. In particular the original deep potential and its phase
equivalent partner must then produce identical angular spectra at
large distances, where the zero-range limit is reached.

The boundary conditions in Eq.(\ref{bou}) with the explicit
expressions for the angular functions in Eqs.(\ref{limit}) and
(\ref{sin}) result then in a system of linear equations for the
coefficients $A_{i}$, i.e.
\begin{equation}
A_{i}\nu \cos \left( \nu \frac{\pi }{2}\right) + 2\sum_{j\neq i}\frac{%
A_{j}\sin \left[ \nu \left( \varphi _{ij}-\frac{\pi }{2}\right) \right] }{%
\sin (2\varphi _{ij})}=-\frac{\rho }{\widetilde{a}_{i}}A_{i}\sin \left( \nu 
\frac{\pi }{2}\right) \;,\;i=1,2,3\;.
\end{equation}
This system of equations has non-trivial solutions only when the
corresponding determinant $D$ is zero, i.e. $D \equiv$ det$\{M_{ij}\}
= 0$ with
\begin{eqnarray} \label{eigen}
 M_{ij} = 2\frac{\sin \left[ \nu \left( \varphi _{ij}-\frac{\pi }{2}\right)
 \right] }{\sin (2\varphi _{ij})}+\left[ \nu \cos \left( \nu \frac{\pi }{2}
 \right) +\frac{\rho }{\widetilde{a}_{i}}\sin \left( \nu \frac{\pi }{2}
 \right) \right] \delta _{ij} \;.  
\end{eqnarray}
When two or three of the scattering lengths are infinitely large $D=0$
has a specific negative solution $\nu$ which results in the so-called
Efimov effect \cite {fed93}.

For finite scattering lengths there are two types of solutions to
$D=0$ in the large-distance limit where $\rho \gg \max (\left|
\widetilde {a}_{i} \right| )$. One type of solution is only present
when there is a bound state with binding energy $ E_{b}$ in one of the
two-body subsystems $j$. This is most easily seen if we assume that
$\widetilde{a}_{j} < 0$ and $\left| \widetilde{a}_{j}\right| \gg
R_{j}$ which imply that the binding energy $E_{b}\approx \hbar^{2} /
(2\mu _{i}a_{j}^{2}) \equiv \hbar^{2} /
(2m\widetilde{a}_{j}^{2})$. Then $D=0$ has a solution $\nu$ which
diverges linearly with $\rho$, i.e.
\begin{equation} \label{nubound}
 \nu =-i\frac{\rho }{\widetilde{a}_{i}} \;\;,\;\;
 \lambda = \nu^2 -4 \approx  -\frac{2m\rho ^{2}}{\hbar ^{2}} E_{b} \;.  
\end{equation}
Thus this angular eigenvalue diverges parabolically with $\rho$ at
large distance.

The other type of solutions has eigenvalues $\nu $ approaching
constants as $\rho \rightarrow \infty $. This is understood from the
dominance of the terms proportional to $\rho$ in Eq.(\ref{eigen})
resulting in the solutions
\begin{equation} \label{nu2}
\nu \rightarrow {\rm Const} \;\;,\;\; \sin \left( \nu \frac{\pi }{2}\right) =0
 \;\;,\;\; \nu_{n}=2(n+1)\;,\;n=0,1,...\;.  
\end{equation}
displaying a spectrum identical to that obtained for small $\rho
$. These eigenvalues $\nu$ can be found to next order in an expansion
in powers of $\rho $ by expanding $D$ as
\begin{eqnarray}
  D = \det\{M_{ij}\} &=& \rho ^{3}
\frac{1}{\widetilde{a}_{1}\widetilde{a}_{2}\widetilde{a}_{3}}
\sin ^{3}\left( \nu \frac{\pi }{2}\right)\\ \nonumber
&+&\rho ^{2}\left(
\frac{\widetilde{a}_{1}+\widetilde{a}_{2}+\widetilde{a}_{3}}
{\widetilde{a}_{1}\widetilde{a}_{2}\widetilde{a}_{3}}
\right) \nu \cos \left(
\nu \frac{\pi }{2}\right) \sin ^{2}\left( \nu \frac{\pi }{2}\right) 
+O(\rho)\;,
\end{eqnarray}
where $O(\rho)$ at most increases as $\rho$ for $\rho \rightarrow
\infty$. These leading two terms then give the eigenvalue equation
$D=0$, i.e.
\begin{equation} \label{eigen-a}
 \sin \left( \nu \frac{\pi }{2}\right) +
 \nu  \cos \left( \nu \frac{\pi }{2}\right)  \frac{1}{\rho }
 \sum_{i}\widetilde{a}_{i}  =0  \;,  
\end{equation}
which has an explicit solution for the inverse function $\rho (\nu )$,
i.e.
\begin{equation}
 \rho (\nu )= - \frac{\sin \left( \nu \frac{\pi }{2}\right) }
 {\nu \cos \left( \nu \frac{\pi }{2}\right) }  \sum_{i}\widetilde{a}_{i}  \;.
\end{equation}
The eigenvalue solutions $\nu_n$ to Eq.(\ref{eigen-a}) are then to
first order in $\rho ^{-1}$ given by
\begin{eqnarray}
 \nu _{n} &=&2(n+1)-\frac{4(n+1)}{\pi }\frac{\sum_i \widetilde{a}_i}{\rho } 
 \;, \\ \lambda _{n} &=&\nu _{n}^{2}-4=(2(n+1))^{2}-4-
 \frac{16(n+1)^{2}}{\pi } 
 \frac{ \sum_i \widetilde{a}_i}{\rho } \;, \label{large}
\end{eqnarray}
where the leading terms already are given in Eq.(\ref{nu2}).

In conclusion, the phase equivalent potential has at large distances
the eigenvalue spectrum in Eqs.(\ref{nu2}) and (\ref{large}) while the
original deep potential has the same spectrum plus the additional
eigenvalue in Eq.(\ref{nubound}) resulting from the bound state.  More
two-body bound states would simply result in precisely the same number
of parabolically diverging eigenvalues of the form in Eq.(\ref{nu2}).
Therefore except for this (or these) additional eigenvalue(s) the angular
spectra of the two potentials are asymptotically identical in the
large-distance limit.

\subsection{Connecting small and large-distence eigenvalues}

The original and the phase equivalent potentials have identical
angular eigenvalue spectra at both large and small distances when only
one two-body subsystem has bound states. This agreement includes at
least the first two terms in the expansion both for small and large
$\rho$. The deep potential has in a one-to-one correspondence with the
two-body bound states in addition a number of low-lying eigenvalues,
which diverge parabolically towards $-\infty$ for $\rho \rightarrow
\infty$. The structure of the related states is asymptotically
precisely that of the bound two-body states with the third particle in
the continuum. When more than one two-body subsystem has bound states
the small-distance behavior of the two spectra is only approximately
identical.

We can now formulate a convincing prescription for dealing with Pauli
forbidden states in three-body cluster computations. In complete
analogy with the two-body problem, where the lowest eigenvalues of the
two-body hamiltonian are forbidden, we declare the lowest angular
eigenvalues, or equivalently the lowest adiabatic potentials, as Pauli
forbidden. We consequently remove them from the basis for all values
of $\rho$ in the subsequent solutions of the radial equations. The
resulting spectra for the original potential with the Pauli forbidden
states and its phase equivalent partner are then approximately
identical for all values of the hyperradius.

\begin{figure} [ht]
\centerline{\psfig{file=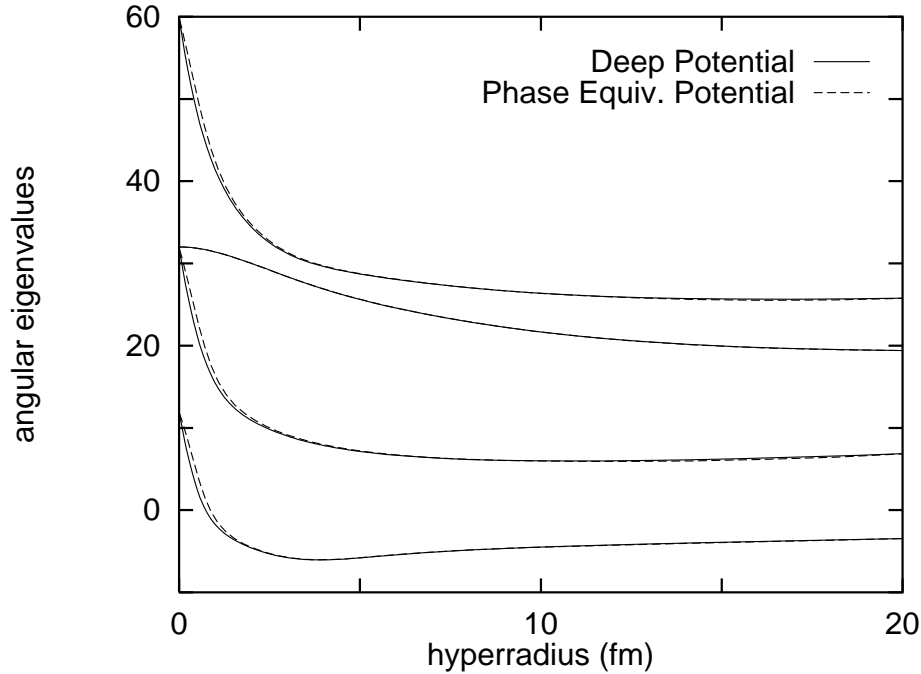,width=12cm,%
bbllx=2.2cm,bblly=1.7cm,bburx=16.5cm,bbury=12.2cm}}
\vspace{0.5cm}
\caption{The lowest angular eigenvalues $\lambda_n$ as functions of
hyperradius for the original schematic potential and its phase
equivalent partner, see Eqs.(\ref{pep0}) and (\ref{dpep0}). The
parameters are $\alpha=4.151$fm$^{-1}$ and $\beta=1.063$ fm$^{-1}$,
$\hbar^2/(2\mu) = 41.4$ MeV fm$^2$ and $m_3 = 18 \mu$. The other two
potentials are equal and given by $V_{13}=V_{23}=-7.8$ MeV
$\exp(-(r/2.55 {\rm fm})^2)$. The lowest parabolically diverging level
corresponding to the Pauli forbidden state has been removed. Only
$s$-waves are included in the computation.}
\label{fig1}
\end{figure}

The small differences must be compensated by slightly different
three-body interactions, see Eq.(\ref{rad1}). This adjustment is
necessary in any case to ensure that the three-body binding energy is
sufficently precise. Even two-body interactions reproducing the
two-body phase shifts would otherwise underestimate the three-body
binding energy. The three-body potential is related to polarization of
the particles beyond that included already in the two-body
potentials. It must then arise when all three particles interact
simultaneously, i.e. are close to each other. The three-body potential
may also be viewed as an adjustment of the off-shell behavior of the
two-body potentials exclusively determined by the on-shell behavior.
After the fine tuning practically indistinguishable and almost exactly
identical results are obtained by use of the two different two-body
potentials with different appropriate three-body potentials.

\section{Numerical illustration}

The three-body angular eigenvalue spectra are identical or at least
approximately identical for short distances for phase equivalent
two-body potentials. In addition, the large-distance behavior is
determined by the scattering lengths and therefore also identical for
such potentials. The intermediate distances must be studied
numerically.

\begin{figure}[ht]
\centerline{\psfig{figure=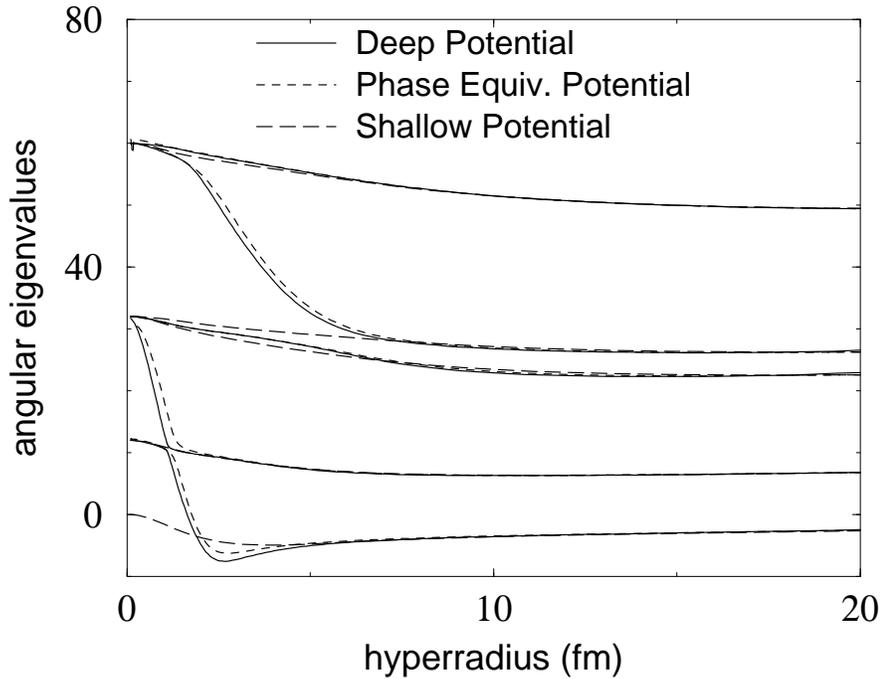,width=12cm,%
bbllx=3.3cm,bblly=2.0cm,bburx=18.7cm,bbury=21.7cm,angle=270}}
\vspace{0.2cm}
\caption[]{The lowest three-body angular eigenvalues $\lambda_n$ as
functions of hyperradius for realistic $s$-wave potentials for the
$^9$Li+n+n system. The neutron-neutron interaction is from
\cite{gar97}. We use three different neutron-$^9$Li potential with the
same low-energy scattering properties, i.e. (i) a deep potential with
Pauli forbidden states, $V_{nc} = - 288$ MeV $\exp(-(r/ 1.15 {\rm fm})^2)$,
(ii) its phase equivalent partner without bound states and (iii) a
shallow potential without bound states, $V_{nc} = -5.6$ MeV
$\exp(-(r/2.55{\rm fm})^2)$ from \cite{gar98}.  Only levels
antisymmetric with respect to interchange of the two neutrons are
shown. We also removed the lowest parabolically diverging level
corresponding to the Pauli forbidden state. Only $s$-waves are
included in the computation.}
\label{fig2}
\end{figure}

We shall first consider a three-body system where only one two-body
subsystem has a bound state. We choose the schematic potentials in
Eqs.(\ref{pep0}) and (\ref{dpep0}) for these two particles. The
parameters are chosen to reproduce the neutron-neutron scattering
length of 18.45 fm and the effective range of 2.83 fm. The two
remaining interactions with the third particle are assumed to be equal
and given by $V_{13}=V_{23}=-7.8$ MeV $\exp(-(r/2.55{\rm fm})^2)$
\cite{joh90}. The masses correspond to two neutrons and $^{9}$Li.  The
three-body system then resembles the halo nucleus $^{11}$Li
($^{9}$Li+n+n) except that now the deep neutron-neutron potential has
a bound state.

The resulting lowest angular eigenvalues are shown in Fig. \ref{fig1}
when only $s$-waves are included in the calculation. The lowest
diverging eigenvalue of the original deep potential with the Pauli
forbidden state is removed. The two-body potentials behave as $r^{-1}$
at the origin and the eigenvalues are therefore according to
Eq.(\ref{small}) linear functions of $\rho$ for small distances around
zero. The hyperspherical spectrum is obtained at large distances. The
two spectra are almost indistinguishable for all distances from zero
to infinity.

When more than one two-body subsystem has bound states the
short-distance behavior of the angular spectra for phase equivalent
potentials are only approximately equal. We show in Fig. \ref{fig2}
numerical examples for realistic $s$-wave potentials for three-body
calculations of $^{9}$Li+n+n, where we assume spin-zero for the
$^{9}$Li-core. With three-body interactions for fine tuning these
potentials all produce a $^{11}$Li structure in agreement with
measurements. By construction the $s$-wave phase shifts for the deep
and phase equivalent potentials are identical for all energies, and
indistinguishible from those of the shallow potential for energies
below 15 MeV.

\begin{figure}[ht]
\centerline{\psfig{figure=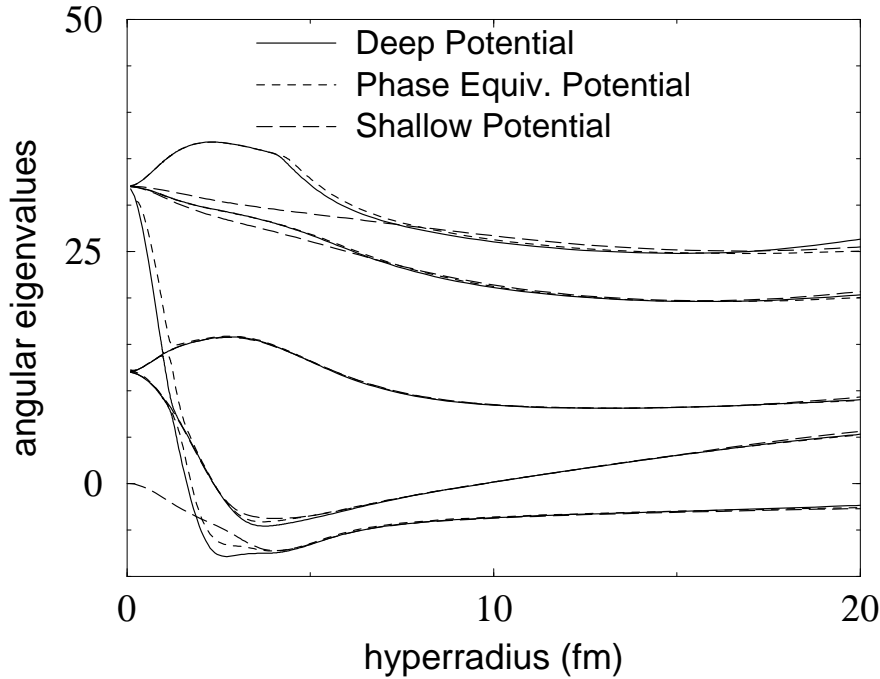,width=12cm,%
bbllx=3.3cm,bblly=2.0cm,bburx=18.7cm,bbury=21.7cm,angle=270}}
\vspace{0.2cm}
\caption[]{The same as Fig. \ref{fig2}, but now we included both $s$
and $p$-waves in the neutron-$^{9}$Li interaction. The Pauli forbidden
states are two-body $s$-waves and the interactions for $s$-waves are
as in Fig. 2. For $p$-waves we use the interaction in \cite{gar98}
denoted potential IV, with a central interaction given by $-5.0$ MeV
$\exp(-(r/2.55 {\rm fm})^2)$, and a spin-orbit interaction given by
$33.6$ MeV ${\bf {l}}_{nc} \cdot {\bf {s}}_{n} \exp(-(r/2.55 {\rm
fm})^2)$.  These interactions are realistic for use in calculations of
halo properties.}
\label{fig3}
\end{figure}

The two spectra arising from the truly phase equivalent potentials,
(i) and (ii) in Fig. \ref{fig2}, are again almost indistinguishable
although they differ slightly more at small and intermediate distances
than those in Fig. \ref{fig1}. Otherwise the features of the schematic
potential remain for these two potentials. On the other hand the
spectrum from potential (iii) differs qualitatively from the other two
spectra in spite of the fact that the low-lying $s$-wave phase shifts
are identical for all these three potentials. The hyperspherical
spectrum is still for the shallow potential obtained in both limits of
$\rho = 0$ and $\rho = \infty$, but now the lowest level starts at
$\lambda = 0$ like the (removed) Pauli forbidden level for the deep
potential.  The phase equivalent potential does not have any level at
$\lambda = 0$.  The large-distance behavior coincide precisely for all
three potentials. The angular eigenvalues are in general more smooth
for the shallow potential.

The different behavior at short distance must of course be reflected
in different phase shifts at higher energies, but low-energy
properties are only marginally influenced. For the same reason the
spatially extended halo structure resulting from these three
potentials can hardly be distinguished. This may also be expressed as
a dominance of intermediate and large distances in the radial wave
function. 

The $s$-wave potentials used in Fig. \ref{fig2} are realistic choices
for computations of the three-body properties of $^{11}$Li. The
$s$-waves are dominating for this system, but the contributions from
$p$-waves are also necessary. These partial waves couple in the
three-body computation and several new angular eigenvalues appear as
seen in Fig. \ref{fig3}. Comparing to Fig. \ref{fig2} we can identify
the new levels which predominantly are of $p$-wave character. Even
with these couplings the quantitative features of the three potentials
are still remarkably similar with the exception of the lowest level
for the repulsive $s$-wave potential.

\begin{figure}[ht]
\centerline{\psfig{figure=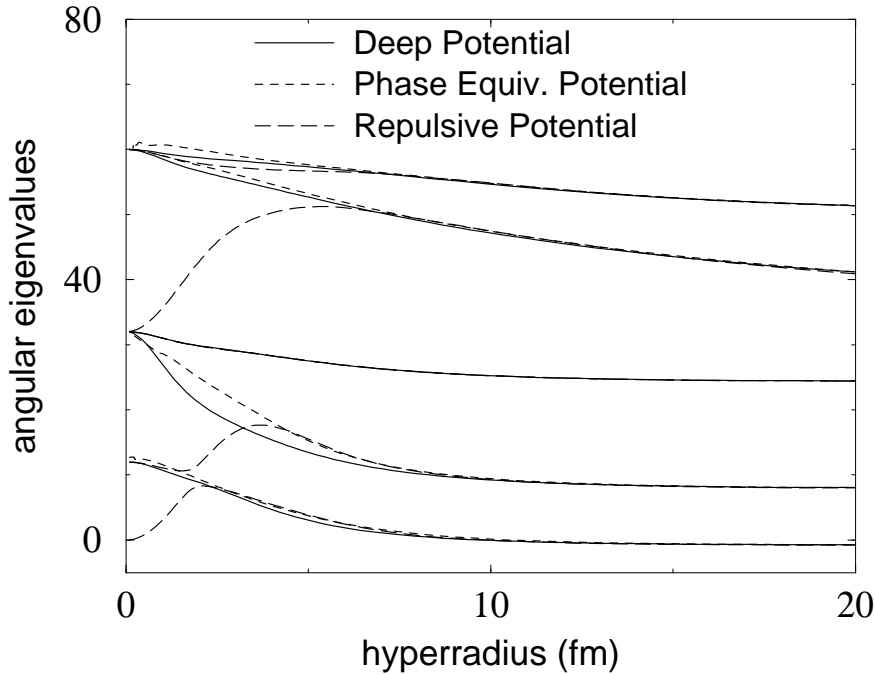,width=12cm,%
bbllx=3.3cm,bblly=1.9cm,bburx=18.6cm,bbury=21.7cm,angle=270}}
\vspace{0.2cm}
\caption[]{The same as Fig. \ref{fig2} for the $^4$He+n+n system. The
deep potential is given by $-75.06$ MeV $\exp(-(r/1.43 {\rm fm})^2)$ and
now the shallow potential is substituted by the repulsive potential 48
MeV $\exp(-(r/2.33{\rm fm})^2)$. The measured low-energy $s$-wave
phase shifts are reproduced by all potentials.}
\label{fig4}
\end{figure}

We also want to compare the spectra when the $p$-waves dominate in the
three-body structure while the Pauli forbidden state still remains in
one of the two-body $s$-wave potentials. A halo nucleus with these
features is $^6$He ($^4$He+n+n) where the shallow $s$-wave potential
for $^{11}$Li is substituted by a realistic repulsion. All three
potentials (deep, phase equivalent, repulsive) are adjusted to
reproduce the measured low-energy $s$-wave phase shifts. All three
angular eigenvalue spectra are shown in Fig. \ref{fig4} when only
$s$-waves are included.

The repulsion is revealed at small distance by the increasing levels
which relatively fast join the corresponding levels from the other two
potentials. For the shallow $s$-wave potential in Fig. \ref{fig2} the
levels from the two phase equivalent potentials varied in contrast
relatively fast at small distances, but the outcome is in both cases
that all spectra quickly becomes very similar and at large distances
completely indistinguishable. The overall quantitative agreement
between these spectra is again remarkably good.

\begin{figure}[ht]
\centerline{\psfig{figure=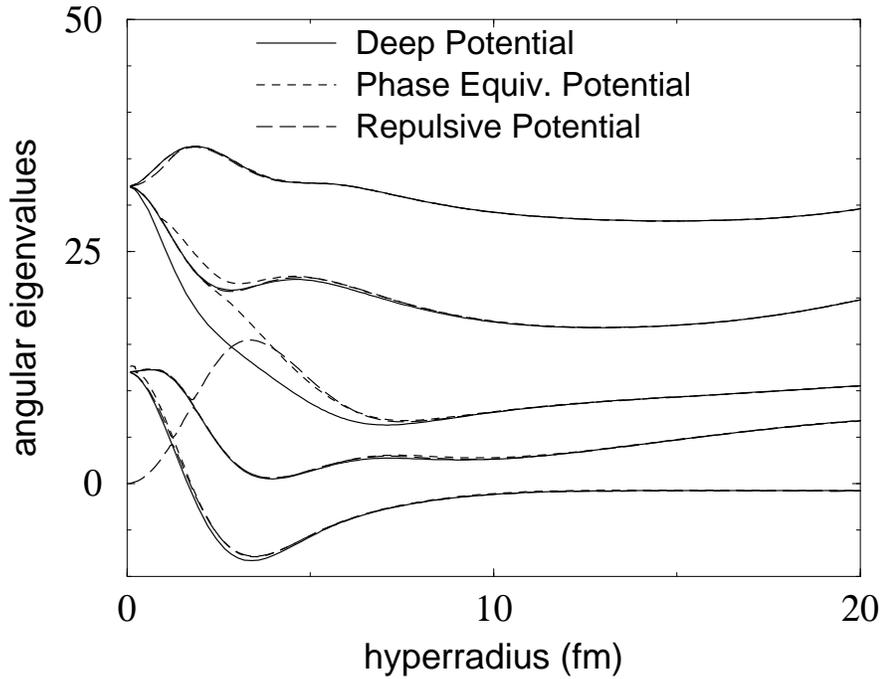,width=12cm,%
bbllx=3.3cm,bblly=2.0cm,bburx=18.7cm,bbury=21.7cm,angle=270}}
\vspace{0.2cm}
\caption[]{The same as Fig. \ref{fig4}, but now we included both $s$,
$p$ and $d$-waves. The $s$-potentials are as in Fig. \ref{fig4}, while
the $p$, $d$ and spin-orbit interactions are taken from \cite{cob98}.
The measured low-energy $s$-wave phase shifts for all partial waves
are reproduced by all potentials. The Pauli forbidden states are
two-body $s$-waves.}
\label{fig5}
\end{figure}

Including $s$, $p$ and $d$-waves in the computation produce more
levels and a number of avoided crossings as seen in Fig. \ref{fig5}.
However, the larger variation with hyperradius does not spoil the
close quantitative resemblance of the three spectra. Instead the
agreement is actually improved due to the coupling of the different
partial waves and the related avoided crossings. For example the
$s$-wave increasing from the starting point of $\lambda = 32$ on
Fig. \ref{fig4} is now on Fig. \ref{fig5} quickly joining another
level also originating from 32 but with higher angular momentum. Thus,
the angular spectra are also remarkably similar when the $p$-waves
dominate as in realistic computations of the three-body properties
of the $^6$He system.

\begin{figure}[ht]
\centerline{\psfig{figure=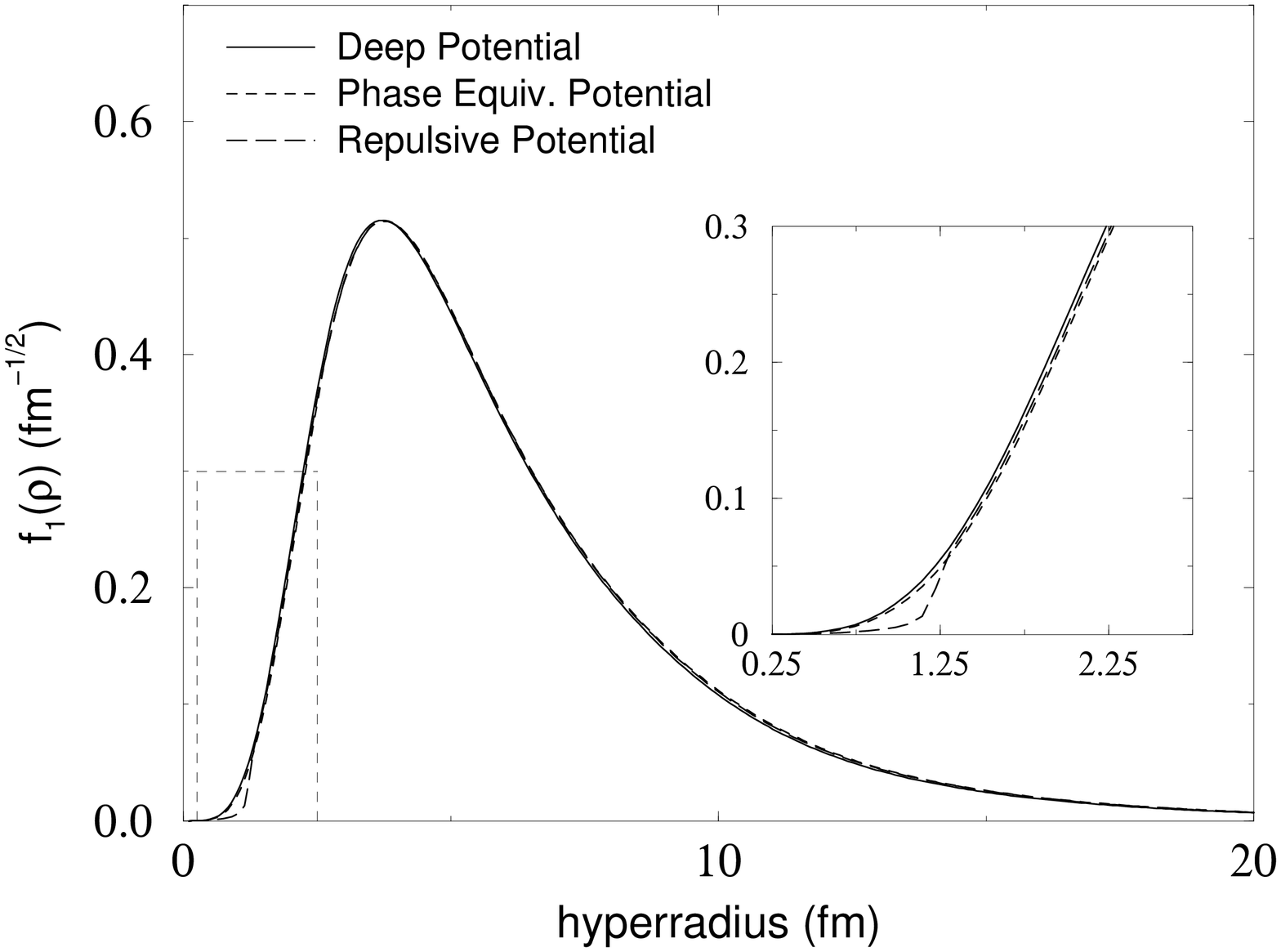,width=12cm,%
bbllx=3.7cm,bblly=1.5cm,bburx=20.0cm,bbury=23.5cm,angle=270}}
\vspace{0.2cm}
\caption[]{The lowest radial wave function as function of hyperradius
for the $^4$He+n+n system for the potentials used in Fig. \ref{fig5}.
The inset shows the small distances where the wave functions differ
from zero and that of the repulsive potential from the two other
potentials.}
\label{fig6}
\end{figure}

The radial wave function is obtained from Eq.(\ref{rad1}) and the
lowest is often carrying more than 95 \% of the probability. To
demonstrate the similarity of the results found by use of the
different potentials it is therefore sufficient to show the lowest
function $f_1$, see Fig. \ref{fig6}. Indeed the quantitative agreement
is overwhelming and the curves are essentially indistinguishable
except perhaps at the very small hyperradii (blown up and shown in the
inset). The slightly smaller probability around the hyperradius of 1
fm for the repulsive potential simply reflects that the effective
potential is larger in this region, see Fig. \ref{fig5}. On the other
hand the probability is already small and the difference is therefore
insignificant. Even these qualitatively different two-body potentials
lead to very similar radial wave functions provided the phase shifts
are identical. The two phase equivalent potentials produce practically
the same radial wave functions for all hyperradii.

\section{Summary and conclusion}

We use the adiabatic expansion and the hyperspherical coordinates to
solve the three-body problem for given short-range two-body
potentials.  When the three particles are clusters consisting of
identical fermions the antisymmetrization between clusters is
technically difficult and a number of different practical
approximations are used. For nuclear halos the particles are often
nucleons interacting with a nucleus. Then this problem can be
formulated as exclusion of Pauli forbidden states within the nucleus
from the space suspending the physical three-body states. In two-body
calculations this is straightforward by direct projection but the use
of phase equivalent potentials still provide an interesting
alternative.

We use phase equivalent two-body potentials with a different number of
bound states and compare the corresponding three-body angular
eigenvalue spectra. We show analytically that these spectra, or
equivalently the adiabatic radial potentials, at large distances are
identical to second order, except for a number of diverging
eigenvalues precisely corresponding to the additional Pauli forbidden
two-body bound states. This limit is mathematically obtained by use of
zero-range potentials and the complete analytical solution is
therefore derived and discussed.

We show furthermore analytically that the spectra at small distances,
with precisely the same exception for the additional bound states,
also are identical to second order provided only one of the two-body
potentials differ in the number of bound states. When more than one of
the two-body potentials differ in the number of bound states the
spectra at small distances are only approximately identical. We give
an estimate in perturbation theory of this relatively small deviation.

In the numerical illustrations we use schematic as well as realistic
potentials for the two-neutron halo nuclei $^{11}$Li ($^9$Li+n+n) and
$^6$He ($^4$He+n+n). The bound states are for convenience assumed to
be $s$-states and the two halo nuclei then provide tests on systems
where respectively the $s$ and $p$-configurations dominate in the
three-body structure. The three-body angular spectra for phase
equivalent potentials are in all cases in remarkable numerical
agreement also at intermediate distances. The radial wave function is
finally computed for the different potentials. They turn out to be
almost indistinguishable for strictly phase equivalent potentials.

However, we find a small difference at very small distances when we
use an approximate two-body potential reproducing only the low-energy
phase shifts without complete phase equivalence. Such phenomenological
potentials have often been used in practical computations of nuclear
halo properties. Due to the large spatial extension of halos the
probability is by definition very small at small distances and the
deviation is therefore insignificant for all practical purposes.

We formulate a prescription to deal with the Pauli principle for
two-nucleon halos. Assume that the two-body potentials have Pauli
forbidden states. We then construct and use the phase equivalent
partners without these forbidden states. The difference between the
three-body adiabatic potentials arising from the original and the
phase equivalent two-body potentials is essentially only additional
basis functions in the hyperspherical adiabatic expansion.  Removing
these from the adiabatic basis makes the resulting adiabatic spectra
asymptotically identical both at small and large distances and
approximately equal as well at all the intermediate distances.

The removed basis functions correspond at large distances precisely to
two particles in Pauli forbidden states and the third particle in a
continuum state. At small distances the lowest levels of the original
deep potentials correspond to identical fermions occupying the same
states. Removal of these terms therefore forces the particles to
occupy higher-lying orbits and thereby introducing the necessary
repulsion preventing violation of the Pauli principle. In conclusion,
this method to exclude the Pauli forbidden states in a three-body
system has firm mathematical and numerical foundations. It is a
practical and accurate alternative to the other existing methods.

\end{document}